\documentclass[aps,showpacs]{revtex4}
\usepackage{graphicx}
\usepackage{amssymb}
\usepackage{amsmath}
\newcommand{\be}{\begin{equation}}
\newcommand{\ee}{\end{equation}}
\newcommand{\ben}{\begin{eqnarray}}
\newcommand{\een}{\end{eqnarray}}
\newcommand{\bes}{\begin{subequations}}
\newcommand{\ees}{\end{subequations}}
\newcommand{\bb}{\bibitem}

\begin{document}
\title{First-order formalism and dark energy}
\author{D. Bazeia, C.B. Gomes, L. Losano, and R. Menezes}
 \affiliation{Departamento de F\'\i sica, Universidade Federal da Para\'\i ba\\
 Caixa Postal 5008, 58051-970 Jo\~ao Pessoa, Para\'\i ba, Brazil}
\date{\today}

\begin{abstract}
This work deals with cosmological models driven by real scalar field, described by standard dynamics in generic spherical, flat, and hyperbolic geometries. We introduce a first-order formalism, which shows how to relate the potential that specifies the scalar field model to Hubble's parameter in a simple and direct manner. Extensions to tachyonic dynamics, and to two or more real scalar fields are also presented.
\end{abstract}


\maketitle

%%%%%%%%%%%%%%%%%%%%%%%%%%%%%
%%%%%%%%%%%%%%%%%%%%%%%%%%%%%
\section{Introduction} 

The recent discovery of cosmic acceleration \cite{a1,a2,a3} has deeply affected modern cosmology. One of the main mysteries of the cosmic acceleration is the so-called dark energy, which makes up a very significant portion of the total energy of the universe. The simplest way to deal with dark energy includes a cosmological constant, but we can also consider standard Friedmann-Robertson-Walker (FRW) models described by real scalar fields, as in quintessence \cite{q1,q2,q3,q4}. For the interested reader we recommend Refs.~{\cite{prr,pr,tc,li}} for some recent reviews on the subject.

In this Letter we focus on dark energy, that is, we turn attention to FRW models described by real scalar fields. Our investigations inspect Einstein's equation and the equation of motion for the scalar field in a very direct way. We consider models described by real scalar field in generic space-time, displaying the usual spheric, flat or hyperbolic spatial profile. However, we follow a very specific route, in which we use the potential of the scalar field to infer how the scale factor or, alternatively, Hubble's parameter evolves in time.

The power of the method that we develop in this Letter is related to an important simplification, which leads to models governed by scalar field potential of very specific form, depending on a new function, $W=W(\phi),$ usually named superpotential when supersymmetry is present, which very much remind us of supergravity, although we do not deal with supersymmetry in the present work. As we show below, we directly relate the function $W(\phi)$ with Hubble's parameter, and this leads to scenarios of current interest to cosmology, unveiling a new route to investigate the subject. Almost all calculations are done as directly as possible, and we illustrate the main results with examples of current interest to modern cosmology.

Models for FRW cosmology with a single real scalar field are described by the standard action
\be\label{model}
S=\int\,d^4x\;{\sqrt{-g}\;\left(-\frac14\,R+{\cal L(\phi,\partial_\mu\phi)}\right)}
\ee
where $\phi$ describes a real scalar field and we are using ${4\pi G}=1.$ The line element is $ds^2=dt^2-a^2(t)d{\vec r}^{\,2}$, and $a(t)$ is the scale factor.
In general, the energy-momentum tensor is given by $T^\mu_{\;\;\nu}=(\rho,-p,-p,-p),$ where $\rho$ and $p$ represent energy density and pressure.
We use Einstein's equation to get 
\bes\label{fe}
\ben
H^2&=&2\rho/3-k/a^2
\\
{\ddot a}/{a}&=&-(\rho+3p)/3
\een
\ees
where $k$ is constant: $k=1,0,$ or $-1,$ for spherical, flat, or hyperbolic geometry, respectively.

The equation of motion for the scalar field depends on ${\cal L}(\phi,\partial_\mu\phi),$ which has the standard form 
\ben\label{sm}
{\cal L}=\frac12\partial_\mu\phi\partial^\mu\phi-V(\phi)
\een
The energy density and pressure are given by $\rho=\dot\phi^2/2+V$ and $p=\dot\phi^2/2-V,$ and the equation of motion for the scalar field has the form
\be
\ddot\phi+3H\dot\phi+V_{\phi}=0\label{em1}
\ee
where $H=\dot a/a$ stands for Hubble's parameter, and $V_{\phi}$ represents $dV/d\phi.$ 

%%%%%%%%%%%%%%%%%%%%%%%%%%%%%%%%
%%%%%%%%%%%%%%%%%%%%%%%%%%%%%%%%
\section{The procedure}

We now introduce our methodology. We use the energy density, pressure and Eqs.~(\ref{fe}) to get, firstly using $k=0$ for simplicity,
\bes\label{em0}
\ben
{\dot H}=-\dot\phi^2\label{em2}
\\
H^2={\dot\phi}^2/3+2V/3\label{em3}
\een\ees
Thus, the set of Eqs.~(\ref{em1}) and (\ref{em0}) constitutes the equations we have to deal with in the case of scalar field with standard dynamics
in flat geometry.

In the standard view, since $a=a(t)$ and $\phi=\phi(t),$ we have $H=H(t)$ and from Einstein's equation we need to see the potential as a function of time. However, from the equation of motion for the scalar field we have $V=V(\phi);$ thus, to make these two views equivalent, we then need to view Hubble's parameter as a function of the scalar field. This is the key point, and we make it very efficient with the introduction of a new function -- $W=W(\phi)$ -- from which we can now understand that Hubble's parameter depends on time as a function of $W[\phi(t)].$
That is, we write
\be\label{be0}
H=W(\phi)
\ee
This is a first-order differential equation for the scale factor. It allows obtaining another equation, involving the scalar field, in the form
\be\label{be}
\dot\phi=-W_{\phi}
\ee
which is also a first-order equation. The two Eqs.~(\ref{be0}) and (\ref{be}) allow writing the potential as
\be\label{pot}
V=\frac32 W^2-\frac12 W^2_{\phi}
\ee
We now notice that for the above potential, solutions of the two first-order Eqs.~(\ref{be0}) and (\ref{be}) also solve the set of Eqs.~(\ref{em1}) and (\ref{em0}). Also, the deceleration parameter is given by $q=-{\ddot a}a/{\dot a}^2=-1-{\dot H}/H^2,$ and here it has the form $q=-1+(W_\phi/W)^2.$

The above calculation directly leads to the potential in Eq.~(\ref{pot}), which very much reminds us of supergravity, where supersymmetry imposes similar restriction \cite{cgr,st,dfgk,bcy,bbn}. However, our calculation has nothing to do with supergravity, which requires more sophisticated manipulations. By the way, we notice that the potential does not depend on the sign of $W,$ and so the change $W\to-W$ in the above equations in general leads to another possibility: $H=-W$ and $\dot\phi=W_\phi.$ A careful search on the subject has led us to a work of Kallosh and Linde \cite{kl}, in which they pointed out similar possibilities, although looking to the problem from another point of view, from a supersymmetric braneworld perspective.

Much of the above methodology was inspired on former works, in which one uses flat space-time to investigate first-order equations for two or more real scalar fields \cite{bds}, up to arbitrary dimensions \cite{bmm}. The positive answer motivates the inspection of the case with generic $k.$ Here the set of equations is given by Eq.~(\ref{em1}) and 
\bes\label{emk}
\ben
\dot H=-\dot\phi^2+k/a^2\label{emk2}
\\
H^2={\dot\phi}^2/3+2V/3-k/a^2
\een\ees
This case is harder to implement, and we notice that the use of $H=W$ does not suffice to solve the problem anymore. However, if we insist with $H=W,$ the procedure now requires the presence of a new constraint. We follow this line, and we suppose that
\be\label{bek}
\dot\phi=\left(k\alpha Z-W_\phi\right)
\ee
where $Z=Z(\phi)$ is in principle arbitrary function and $\alpha$ is constant. This Eq.~(\ref{bek}) is similar to the former Eq.~(\ref{be}), valid in the case $k=0,$ but now the potential has to be changed to the form
\be
V=\frac32W^2+\left(k\alpha Z-W_\phi\right)\left(k\alpha Z+\frac12W_\phi\right)
\ee
The constraint emerges consistently; it can be written as
\be\label{cons1}
W_{\phi\phi}Z+W_\phi Z_\phi-2k\alpha Z Z_\phi-2W Z=0
\ee 
and this opens several possibilities, as we discuss below.

We illustrate the above results with some examples. Firstly, we consider the case $k=0.$ We take $W=A\phi^n,$ $A$ and $n$ constants. This gives the potential 
\be\label{potnew}
V=\frac12 A^2\phi^{2n}\left(3-\frac{n^2}{\phi^2}\right)
\ee
and Hubble's parameter is $H=A^{\frac2{2-n}}\;[n(n-2)\,t]^{\frac{n}{2-n}}.$ This expression excludes the cases $n=1$ and $2,$ which have to be investigated separately. The case $n=2$ gives $H=A e^{-4 A t},$ and now the potential has the form $V(\phi)=(A^2/2)(3\phi^2-4)\phi^2,$ which presents spontaneous symmetry breaking. The case $n=1$ is also interesting; it reproduces some of the `negative' potentials investigated in Ref.~{\cite{np}}.

Another example is given by $W=A e^{-B \phi},$ $A$ and $B$ constants. This gives 
\be
V=\frac12 A^2(3-B^2)e^{-2B\phi}
\ee
with $H=1/B^2 t$ and $q=-1+1/B^2.$

The case with $k\neq0$ is more involved. We see from Eq.~(\ref{cons1}) that there are distinct cases to be considered. For $Z=1$ we get $W_{\phi\phi}=2W,$ and so $W=Ae^{\pm\sqrt{2}\phi}.$ This case leads to diverse decelerations, depending on the values of $A, k$ and $\alpha.$
Another possibility is $Z=W_\phi,$ and this gives $(1-k\alpha)W_{\phi\phi}=W,$ which shows that $W$ is now bounded or not, depending on the sign of $1-k\alpha.$ The deceleration parameter in this last case is given by $q=-1+(1-k\alpha)(W_\phi/W)^2,$ which can be lower or greater than $-1,$ for $1-k\alpha$ negative or positive, respectively. As an example, we choose $W=A e^{B\phi},$ $A$ constant,  $B=\pm1/\sqrt{1-k\alpha},$ and $\alpha<0.$ We get
\be
V=(1-k\alpha)\,A^2\,e^{2B\phi}
\ee
and $H=1/t.$ Here the deceleration parameter vanishes.

We can also consider $W=A\sin(\phi/\sqrt{k\alpha-1})$ for $A$ constant and $k\alpha-1$ positive. In this case we have
\be
V=\frac32 A^2+A^2(k\alpha-1)\cos^2(\phi/\sqrt{k\alpha-1})
\ee
and $\!H\!\!=\!\!A\tanh(At).\!$ Evolution is accelerated, with $q\!\leq\!-1.$

We introduce other examples, to further stress that the above methodology is very efficient. We consider $k=0,$ the case of flat space-time, and we suppose that $W=A[B+\arctan(\sinh(C\phi))],$ $A,$ $B$ and $C$ constants. This gives the potential
\be\label{pc}
V=\frac32 A^2[B+\arctan(\sinh(C\phi))]^2-\frac12 A^2C^2{\rm sech^2}(C\phi)
\ee
We plot this potential for some values of $A,$ $B,$ and $C$ in Fig.~[1]. We illustrate some possibilities with $A=1,$ $C=\sqrt{2}$ and: $B=0,$ which gives a symmetric potential, similar to the form required in the bicycling scenario of Ref.~{\cite{np}}; $B\in(0,\pi/2)$ which gives asymmetric potentials, similar to the case required for cyclic evolution \cite{stu,kst}; $B=\pi/2,$ which gives a kink-like potential, which drives the model to the case of static universe.
%%%%%%%%%%%%%%%%%%%%%%%%%%%%%%%%%%%%%%%%%%%%%%
\begin{figure}[!ht]
\centering
\includegraphics[width=.40\textwidth]{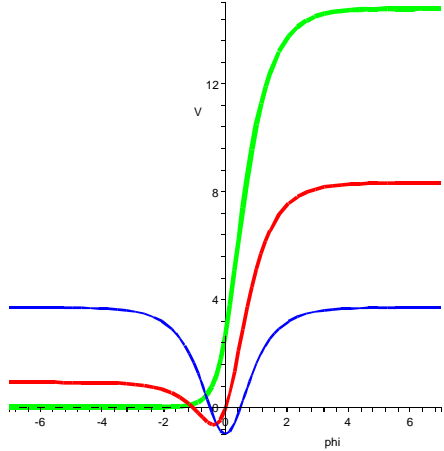}
\caption{The potential of Eq.~(\ref{pc}) for $A=1,$ $C=\sqrt{2},$ and $B=0,\pi/4,$ and $\pi/2,$ plotted with thin, thick, and thicker lines, respectively.}
\end{figure}
%%%%%%%%%%%%%%%%%%%%%%%%%%%%%%%%%%%%%%%%%%%%%%%

We now investigate Hubble's parameter for the above model. The calculation is easy, direct; it allows writing $H=A[B-\arctan(AC^2 t)],$ which shows that $A,$ $B$ and $C$ control the way $H$ evolves in time.

Another example in flat space-time is given by \cite{bmm}
\be\label{wp}
W_p=C_p-\frac{p}{2p-1}\phi^{\frac{2p-1}{p}}+\frac{p}{2p+1}\phi^{\frac{2p+1}{p}}
\ee
where $C_p=2p/(4p^2-1)$ and $p=1,3,...,$ is odd integer. The potential is awkward, but now Hubble's parameter is given by
\be\label{hp}
H_p=C_p-\frac{p}{2p-1}\tanh^{2p-1}(t/p)+\frac{p}{2p+1}\tanh^{2p+1}(t/p)
\ee
which is plotted in Fig.~[2] for some values of $p.$
%%%%%%%%%%%%%%%%%%%%%%%%%%%%%%%%%%%%%%%%%%%%%
\begin{figure}[!h]
\centering
\includegraphics[width=.40\textwidth]{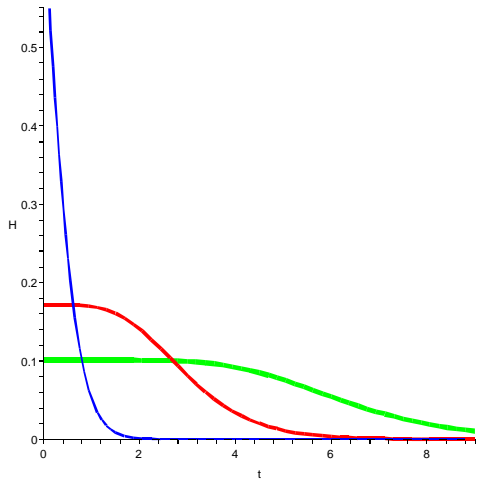}
\caption{Hubble's parameter of Eq.~(\ref{hp}) for $p=1,3,$ and $5,$ plotted with thin, thick, and thicker lines, respectively.}
\end{figure}
%%%%%%%%%%%%%%%%%%%%%%%%%%%%%%%%%
%%%%%%%%%%%%%%%%%%%%%%%%%%%%%%%%
%%%%%%%%%%%%%%%%%%%%%%%%%%%%%%%%

\section{Tachyonic dynamics}

The power of the method can be further extended to the case of scalar field with tachyonic dynamics; see for instance
Refs.~{\cite{t1,t2,t3,t4}} for specific details on scalar fields with tachyonic dynamics, related to the subject under investigation. The Lagrange density has the form
\ben
{\cal L}_t=-V(\phi)\sqrt{1-\partial_\mu\phi\partial^\mu\phi\,}
\een\label{Lt}
The energy density and pressure are given by $\rho_t=V/(1-\dot{\phi}^2)^{1/2}$ and $p_t=-V(1-\dot\phi^2)^{1/2}.$ We consider $k=0,$ and we use Eqs.~(\ref{fe}), the energy density and pressure, and the equation of motion for the scalar field to get
\bes\ben
\ddot\phi+(1-{\dot\phi}^2)(3H\dot\phi+V_\phi/V)=0
\\
\dot H=-V{\dot\phi}^2/(1-{\dot\phi}^2)^{1/2}
\\
H^2=2V/3(1-{\dot\phi}^2)^{1/2}
\een\ees

The case of tachyonic dynamics is more involved, but we keep using $H=W.$ This leads to the equation
\be\label{phi3}
\dot\phi=-\frac23\frac{W_\phi}{W^2}
\ee
and the potential is now given by
\be\label{pott}
V=\frac32\sqrt{W^4-\frac49 W^2_{\phi}}
\ee
Here the deceleration parameter is $q=-1+2W^2_\phi/3W^4.$

We consider an example, in which $W=A/\phi.$ We use Eq.~(\ref{pott}) to obtain
\be\label{pot4}
V=\frac32\,A\sqrt{A^2-\frac49}\,\frac1{\phi^2}
\ee
which requires that $A>2/3.$ We use this result to obtain $H=3A^2/2t,$ and now $q=-1+2/3A^2.$ This example displays the inverse square potential, which is required for the presence of scaling solutions; see \cite{t3} and references therein for other details.

%%%%%%%%%%%%%%%%%%%%%%%%%%%%%%%%%%%%%%%%%
%%%%%%%%%%%%%%%%%%%%%%%%%%%%%%%%%%%%%%%%%
\section{Two-field models}

We now extend the above procedure to the case of two or more real scalar fields with standard dynamics. We consider two-field models for simplicity, and we investigate the important case of flat geometry. Here we have to change the Lagrange density in Eq.~(\ref{sm}) to the form
\be
{\cal L}_2=\frac12\partial_\mu\phi\partial^\mu\phi+\frac12\partial_\mu\chi\partial^\mu\chi-V(\phi,\chi)
\ee
and now we get the new set of equations
\bes\label{em2f}
\ben
\ddot\phi+3H\dot\phi+V_\phi=0
\\
\ddot\chi+3H\dot\chi+V_\chi=0
\\
{\dot H}=-{\dot\phi}^2-{\dot\chi}^2
\\
H^2={\dot\phi}^2/3+{\dot\chi}^2/3+2 V/3
\een\ees
As before, we insist with $H=W,$ but now $W=W(\phi,\chi)$ suggests that we write the two first-order equations
\be
\dot\phi=-W_\phi,\;\;\;\;\;\dot\chi=-W_\chi
\ee 
and the potential
\be
V=\frac32 W^2-\frac12W^2_\phi-\frac12 W^2_\chi
\ee
It is not hard to show that solutions of the above first-order equations also solve the set of Eqs.~(\ref{em2f}). Also, here the deceleration parameter
is $q=-1+(W_\phi/W)^2+(W_\chi/W)^2.$

The above procedure can be generalized to three or more real scalar fields straightforwardly, and it opens interesting possibilities for coupled fields in flat geometry. For instance, if we consider an additive $W,$ that is, if we take $W(\phi,\chi)=W_1(\phi)+W_2(\chi),$ we get the potential in the form $V(\phi,\chi)=V_1(\phi)+V_2(\chi)+3W_1(\phi)W_2(\chi).$ It shows that the interactions appear as the product of the two independent $W_1$ and $W_2.$ Here we see that Hubble's parameter is also additive, that is, $H=H_1+H_2.$ However, acceleration is not additive anymore.

We illustrate the case of two fields firstly with $W$ which involves product of the fields. An example which we solve exactly is given by $W=A\phi+B\chi-C\phi\chi,$ for $A,$ $B,$ and $C$ constants. The potential is
\ben
V&=&-\frac12 (A^2+B^2)+C(A\chi+B\phi)+\frac12(3A^2-C^2)\phi^2\nonumber
\\
& &+\frac12 (3B^2-C^2)\chi^2-3C(A\phi+B\chi)\phi\chi+\frac32C^2\phi^2\chi^2\nonumber
\een
The first-order equations are $\dot\phi=-A+C\chi$ and $\dot\chi=-B+C\phi.$ The solutions require two new constants, $D$ and $E,$
and they give $H=AB/C+CE^2e^{-2Ct}-CD^2e^{2Ct}.$

Another example is for $W$ additive, given by $W=A\phi^2+W_p(\chi),$ $A$ constant, and $W_p$ as in Eq.~(\ref{wp}). The potential is awkward, but
now Hubble's parameter is 
\be\label{h21}
H=A e^{-4At}+H_p
\ee
where $H_p$ is given by Eq.~(\ref{hp}). We plot this parameter in Fig.~[3] to illustrate its behavior in terms of $p,$ for $A=1.$
A comparison between Fig.~[2] and Fig.~[3] shows one field affecting the other behavior explicitly.

Evidently, we can use other models with several fields to get to richer scenarios, as we will explore in another work. Here we learn from the above examples how the scalar fields directly contribute to modify the cosmic evolution, and this is known to be of interest to modern cosmology \cite{rf}.

%%%%%%%%%%%%%%%%%%%%%%%%%%%%%%%%%%%%%%%%%%%%%
\begin{figure}[!h]
\centering
\includegraphics[width=.40\textwidth]{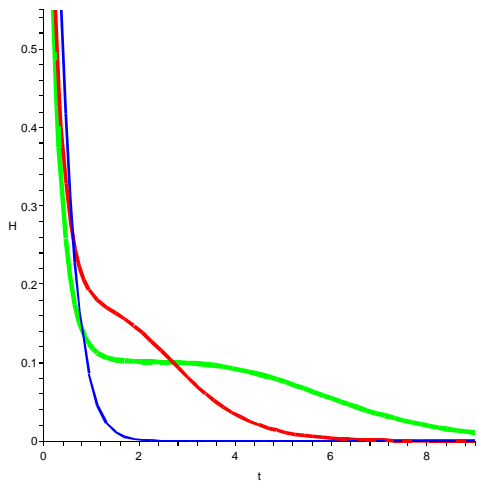}
\caption{Hubble's parameter of Eq.~(\ref{h21}) for $A=1$ and $p=1,3,$ and $5,$ plotted with thin, thick, and thicker lines, respectively.}
\end{figure}
%%%%%%%%%%%%%%%%%%%%%%%%%%%%%%%%%

%%%%%%%%%%%%%%%%%%%%%%%%%%%%%%%%%%%
%%%%%%%%%%%%%%%%%%%%%%%%%%%%%%%%%%%
\section{Ending comments}

In this work we have shown how to write a first-order formalism to FRW cosmology, described by a single real scalar field
with either standard or tachyonic dynamics for generic spherical, flat or hyperbolic spatial geometry. The crucial ingredient was the introduction of a new function, $W=W(\phi),$ from which we could express Hubble's parameter in the form $H(t)=W[\phi(t)].$ The importance of the procedure is related not only to the improvement of the precess of finding explicit solution, but also to the opening of another route, in which we can very fast and directly write $H(t)$ once $W(\phi)$ is given. As we have shown, the present investigations are of direct interest to modern cosmology, since they seem to open several distinct possibilities of investigation. 

The interest in the subject broadens with the extension of the method to the case of several fields, since now we can find models in which one field can be used to affect the behavior of other fields, unveiling the possibility to control a given phase and to link different phases of the cosmic evolution.

The authors would like to thank Francisco Brito and Carlos Pires for discussions, and CAPES, CNPq, PADCT/CNPq, and PRONEX/CNPq/FAPESQ for partial support.

%%%%%%%%%%%%%%%%%%%%%%%%%%%%%%%%%%%%%%%%%%%%%%%%%%%%%%%%%%%%%%%%%%%
%%%%%%%%%%%%%%%%%%%%%%%%%%%%%%%%%%%%%%%%%%%%%%%%%%%%%%%%%%%%%%%%%%%%

\end{document}